\documentclass[preprint,showpacs,preprintnumbers]{revtex4}
\usepackage{amsmath}
\usepackage{latexsym}
\usepackage{graphicx}
\usepackage{graphics}

\input{tcilatex}

\begin{document}

\title{Imperfections in Focal Conic Domains : the Role of Dislocations}
\author{M. Kleman$^{1}$}\email{maurice.kleman@mines.org}
\author {C. Meyer$^{2}$}
\author {Yu. A. Nastishin$^{1,3}$}
\affiliation{$^{1}$Laboratoire de Min\'eralogie-Cristallographie de Paris, CNRS
UMR 7590, T16 case 115, Universit\'e Pierre et Marie Curie, 4 Place Jussieu,
 F-75252 Paris Cedex 05, France\\
$^{2}$Laboratoire de Physique de la Mati\`{e}re Condens\'{e}e, Universit\'{e}
\\
de Picardie, 33 rue Saint-Leu, 80039 Amiens, France\\
$^{3}$Institute of Physical Optics, 23 Dragomanov str. Lviv
79005 Ukraine}
\date{\today }

\begin{abstract}
It is usual to think of Focal Conic Domains (FCD) as perfect
geometric constructions in which the layers are folded into
Dupin cyclides, about an ellipse and a hyperbola that are
conjugate. This ideal picture is often far from reality. We have
investigated in detail  the FCDs in several materials which have
a transition from a smectic A (SmA) to a nematic phase. The
ellipse and the hyperbola are seldom perfect, and the FCD
textures also suffer large transformations (in shape or/and in
nature) when approaching the transition to the nematic phase, or
appear imperfect on cooling from the nematic phase.  We
interpret these imperfections as due to the interaction of FCD's
with dislocations.  We analyze theoretically the general
principles subtending the interaction mechanisms between FCD's
and finite Burgers vector dislocations, namely the formation of
kinks on disclinations, to which dislocations are attached, and
we present models relating to some experimental results. Whereas
the principles of the interactions are very general, their
realizations can differ widely in function of the boundary
conditions.
\end{abstract}

\pacs{61.30Jf, 61.72Lk}
\maketitle

\section{Introduction}
\label{sec:1} The discussion that follows, about the behavior of
defects in the SmA (smectic A) phase is inspired by a few
experimental polarized light microscopy observations reported in
\cite{meyer1} and summarized below.  These observations have
since been developed \cite{yuriy}.  They relate to a domain of
temperature that extends approximately $1^{\circ}C$  below the
SmA $\longrightarrow$  N phase transition, some of the most
relevant experiments having been done with an accuracy of $\pm
1mK$. The very near vicinity of the transition, where phenomena
usually qualified of transitional do happen, is not investigated
here. But, even in the domain we have searched, the focal conic
domains suffer considerable visible modifications, which we
attribute to their interactions with dislocations.  It is
precisely the nature of these interactions that we wish to
describe in the present article.

The defects and textures of the SmA and N phases are reasonably
well understood at mesoscopic and macroscopic scales, at least
for their static physical and topological properties.
Contrariwise, the role played by the smectic defects at the
phase transition has been little investigated.  It is precisely
in this region that the FCD's (focal conic domains), the only
defects that are fully observable in light microscopy, show
large modifications that, we believe, are essentially due to
their interactions with dislocations.  The SmA $\longrightarrow$
N phase transition has been the object of many investigations
(for a review, see \cite{degennes}). The compression modulus $B$
becomes smaller and tends towards a finite value (equal to or
slightly different from zero, see e.g.,
\cite{benzekri,beaubois,yethiraj}); its variation is noticeable
in a large temperature range (more than half a degree in the
compounds that we have investigated).  Note that in this range,
$K_1$ (the splay modulus) stays practically constant. The
question of $\overline{K}$ (the saddle-splay modulus) has been
little investigated yet, either theoretically or experimentally
(see \cite{barbero} for the nematic phase); the results that
follow have been interpreted by assuming that $\overline{K}$ too
stays practically constant. $K_2$ (the twist modulus) is
infinite in layered media, twist being forbidden by the layer
geometry.

Let us now recall some defect features of the SmA phase.  These
defects are of two types, focal conic domains (which are special
types of disclinations) and
dislocations:\\
\noindent - \textit{focal conic domains (FCD's)}: the layers are
parallel, so that there is no strain energy but only curvature
energy.  The normals to the layers envelop two focal surfaces on
which the curvature is infinite (the energy diverges). The focal
surfaces are degenerate into lines in order to minimize this
large curvature energy.  These lines are necessarily two
confocal conics, an ellipse and a hyperbola, observable by
optical microscopy \cite{friedela,friedelb,klemanbook}. The
layers are folded along \textit{Dupin cyclides}, surfaces that
have the topology of tori. And indeed the simplest geometric
case is when the ellipse E is degenerate into a circle - the
confocal hyperbola H being degenerate into a straight line
perpendicular to the plane of the circle and going through its
centre.  In this case the layers are nested tori, restricted in
fact to those parts of the tori that have negative Gaussian
curvature $G=\sigma_1 \sigma_2$. The $G<0$ case is indeed the
most usual case met experimentally in generic Dupin cyclides,
see \cite{maurice,boltenhagen1}. We shall not consider in the
sequel the situations where the layers are restricted to those
parts that have positive Gaussian curvature; and as a matter of
fact the mixed case is not observed. In the toric case just
alluded, the focal conic \textit{domain} is the region of space
occupied by those nested layers restricted to their $G<0$ parts;
it is bound by a cylinder parallel to H and whose cross section
is E. In the generic case, the region of space where the layers
have $G<0$ is bound by two half-cylinders of revolution, that
meet on the ellipse, and whose generatrices are parallel to the
hyperbola asymptotes Fig.1a. This is the picture of an ideal,
\textit{complete}, FCD. Fig.1b illustrates a case where $G<0$
and $G>0$ regions are visible in the same FCD; it does not
correspond to any situation met in practice. Models for
\textit{incomplete} FCD's are shown farther ahead (Fig.9 (a and
b)). The important question how FCD's are packed in space
\cite{friedelb,klemanbook} will be
approached, but just incidentally.\\

The curvature energy $f_{FCD}$ of an entire, ideal, focal conic
domain depends on $K_1$ and $\bar{K}$:

\begin{equation}
f_{FCD}=f_{bulk}+f_{core}=4\pi a (1-e^2)K(e^2)[K_1 \ln
\displaystyle \frac{2b}{\xi}-2K_1-\bar{K}]+f_{core}
\end{equation}
where $a$ is the semi-major axis of the ellipse, $b$ the
semi-minor axis, $e$ the eccentricity and $K(e^2)$ the complete
elliptic integral of the first species \cite{maurice}. It is
believed that the energy $f_{strain}$ attached to the thickness
variation of the layers is negligible compared to $f_{FCD}$.
Very little is known about the core contribution $f_{core}$, but
it is usually assumed that it scales as $a K_1$. Thus, at $a$
and $e$ constant, the FCD total energy does not vary
significantly in the domain of temperature under investigation,
if our assumptions about the temperature variation of $K_1$ and
$\bar{K}$ turn to be
true.\\
\noindent - \textit{screw dislocation lines and edge dislocation
lines:} : their line energies per unit length can be written:\\
\begin{equation}
f_s=\displaystyle \frac{1}{128} \frac{B b_{disl}^4}{r_{c,
screw}^2}+f_{core},  f_{e}=\displaystyle \frac{1}{2} \sqrt{K_1 B}
\displaystyle \frac{b_{disl}^2}{r_{c,edge}}+f_{core},
\end{equation}
where $b_{disl}=n d_0$ is the dislocation Burgers vector ($d_0$
is the layer thickness) and $r_c$ is the core radius. It is
visible that the elastic contributions (the off-core terms in
Eq. 2) decrease when $T$ gets closer to $T_{AN}$, because $B$
decreases and the core energies are expected to be approximately
constant - in a na\"{\i}ve model inspired by the solid-liquid
transition, these energies are of order $k_B(T_{AN}-T)
\displaystyle \frac{\pi r_c^2}{\delta^2 d_0}$ per unit length of
dislocation line, \textit{i.e.} small, $\delta^2 d_0$ being the
volume occupied by a molecule, and $r_c$ being perhaps of order
$\delta$ for a screw dislocation, $d_0$ for an edge dislocation,
\textit{i.e.} practically constant. The decrease of $B$ is
effective on a large temperature domain before the transition,
probably larger than $1^{\circ}C$, see \cite{benzekri,beaubois}.
The core radii scale as the correlation lengths very close to
the transition, but this region is of no interest to us.\\

\noindent \textit{Comments on the experimental conditions}\\

The FCD's are static in the lower range of the domain of
temperature we have investigated, they quite often display
variations to their ideal shape. The transformation of the
FCD's, when approaching the transition, is visible with a simple
optical microscopy set up. It appears as a rather sudden
phenomenon, about half a degree below $T_{AN}$, at a temperature
$T^*$ that depends slightly on the boundary conditions. We shall
assume that it is due to an abrupt multiplication of
dislocations, which then interact with FCD's when they are
mobile enough.\\

The spontaneous multiplication of screw dislocations close to the
SmA $\longrightarrow$ N transition is a well documented fact in
\textit{lyotropic} systems \cite{allaina, allainb, dhez}. Our
observations relate to \textit{thermotropic} compounds, two
belonging to the cyanobiphenyl series, 8CB, 9CB and one belonging
to the cholesteryl series (CN, nonanoate) which also have a Sm A
$\longrightarrow$ N transition; they incline us to believe that
the phenomenon of spontaneous multiplication of dislocations
(screw but also edge) is very general. In these compounds the
focal conic domains suffer considerable modifications in the
region close to the transition. In all cases, either the FCD's
disappear by shrinking before the phase transition, or the ellipse
and the hyperbola transform into disclinations in the nematic
phase; the first situation occurs usually for small and medium
size FCD's slowly heated, the second one occurs for large FCD's,
when they are brought to the transition under faster heating. When
cooling down from the nematic phase, the FCD texture in 8CB, 9CB
and 8OCB usually does not display ideal FCD's. Instead FCD
fragments grow, join and form domains, which in many cases are not
ideal FCD's. The double helical objects described in
\cite{williams1}, which are splitting modes of giant screw
dislocations, are obtained this way. These \textit{imperfect}
FCD's can be quenched to lower temperatures where they stabilize
due to anchoring and viscosity barriers. The boundary conditions
play an important role in the definition of the final texture.

There is little doubt that the transformations of the FCD
texture in 8CB, 9CB, CN, when approaching the nematic phase, as
well as the formation of imperfect FCD's when coming from above,
are due to the interaction of the FCD's with free dislocations.
Dislocations are generally not visible by optical microscopy,
except when their Burgers vector is large (micron size), which
situation occurs for edge dislocations, clustering into oily
streaks \cite{friedela,friedelb,klemanbook} or screw
dislocations split into two $k=\frac{1}{2}$ disclinations
\cite{williams1,kln}. We argue here that the presence of
numerous dislocations can be visualized via
their distorting action on the FCD's, which are visible.\\

\section{GEOMETRIC RULES FOR IDEAL FOCAL CONIC DOMAINS}
Essential for a better understanding of the modifications
suffered by FCD's when interacting with dislocations are the
following properties, that characterize them when they are in an
\textit{ideal} state.

(a)- Projected orthogonally upon a plane, along any direction,
the ellipse E and the hyperbola H cross at right angles, Fig.2a.
This is a particular case of Darboux's theorem \cite{darboux},
which states that if a congruence of straight lines is
orthogonal to a set of parallel surfaces, the two focal surfaces
$\Sigma_1$ and $\Sigma_2$ (that this congruence generically
envelops) are such that the planes tangent to $\Sigma_1$ and
$\Sigma_2$ at the contact points $M_1$ and $M_2$ of any line
$\Delta$ of the congruence are orthogonal. This is the reason
why the projections of the ellipse E and the hyperbola H
belonging to the same FCD look orthogonal. Here the straight
line $\Delta$ is a normal to the Dupin cyclides, and indicates
the average direction of the molecules. Darboux's theorem is
empirically satisfied by a number of FCD's, which in that sense
are ideal FCD's; when it is not, (see Fig.2b) it implies that
the FCD in question is geometrically interacting with other
defects, as we shall discuss in the sequel.

(b)- Two neighboring ideal FCD's whose ellipses are in the same
plane and tangent at some point M are in contact along at least
one line segment joining M to a point P at which the two
hyperbolae intersect. This geometry, frequently observed, is a
particular realization of the law of corresponding cones
\cite{friedela,friedelb,klemanbook}, a geometrical property that
rules the way FCD's pack in space. A tilt grain boundary whose
angle of misorientation $\omega$ is neither too small nor too
large is usually made of a FCD packing such that the ellipses
belong to the grain boundary, have a constant eccentricity $e =
sin \displaystyle \frac{\omega}{2}$, the asymptotes of the
hyperbolae being parallel \cite{kleman1}, see Fig.3.

(c)- in a solid crystal, a tilt grain boundary is usually split
into dislocations. The same is of course possible for a tilt grain
boundary in a layered medium. And indeed the FCD free interstices
(the packing of ellipses in the plane of the grain boundary cannot
be perfect), are filled with dislocations \cite{kleman1}. There is
therefore a relation of equivalence between dislocations and focal
conic domains \cite{bourdon, boltenhagen2}. As a matter of fact,
the ellipse of an isolated FCD is the termination of a set of
dislocations whose total Burgers vector $b_{disl}=4ae=4c$, as
explained below.\\

\section{KINKS ON DISCLINATIONS}
\subsection{Wedge and twist disclinations. FCD confocal conics are disclinations.}

Disclinations are typical line defects in a medium endowed with
a director order parameter \cite{frank}. One distinguishes
\textit{wedge} disclinations, whose rotation vector
$\overrightarrow{\Omega}$ is along the disclination line, and
\textit{twist} disclinations, whose rotation vector
$\overrightarrow{\Omega}$ is orthogonal to the disclination
line. As shown in \cite{friedela,friedelb}, there are
necessarily dislocations attached to a line segment of twist
character.  Let us remind that the focal lines of a FCD, are, by
nature, disclinations.  The hyperbola is a disclination of
strength $k=1$, whose rotation vector
($\overrightarrow{\Omega}=2\pi \overrightarrow{t} $) varies in
direction (not in length) all along the hyperbola: at each point
of the hyperbola it is parallel to the tangent
$\overrightarrow{t}$ at this point. The layer geometry is
axial-symmetric in the vicinity of the hyperbola. Insofar as it
is a disclination, the hyperbola is of wedge character; there
are no attached dislocations.  The ellipse is a disclination of
strength $k=\displaystyle \frac{1}{2}$, whose rotation vector
($\overrightarrow{\Omega}=\pi \overrightarrow{t}$) varies in
direction but (not in length) all along the ellipse;
$\overrightarrow{t}$ is in the plane of the ellipse and tangent
to the layer inside the ellipse, Fig.4. The twist component
induces the
attachment of dislocations \cite{kleman1,boltenhagen1,friedel1}, see below.\\

\subsection{Kinks, generic properties.}

Modifications to the twist/wedge character of a disclination can
be achieved in the generic case by attaching/detaching new
dislocations to the line. Such operations modify the shape of
the line, by the introduction of \textit{kinks}, Fig.5. For
instance, in order to attach at some point A on a wedge line
$\cal L$ a set of dislocations of total Burgers vector
$\overrightarrow{b}_{disl}$, one has to introduce a kink
$\overrightarrow{AB}$, with a component perpendicular to $\cal
L$, (\textit{i.e.} a segment $\overrightarrow{AB}$ having a
twist component), such that
\begin{equation}
\overrightarrow{b}_{disl}=2 \sin \displaystyle \frac {\Omega}{2}
\overrightarrow{t} \times \overrightarrow{AB},
\end{equation}
where $\overrightarrow{t}$ is an unit vector tangent to the line and
$\overrightarrow{\Omega}$ ($\overrightarrow{\Omega}=\Omega \overrightarrow{t}$)
is the rotation invariant carried by the disclination; see
\cite{klemanbook,friedel1} and appendix for a demonstration of Eq. (3). In
practice lines of interest are of strength $|k|=\displaystyle \frac{1}{2}$;
$|\overrightarrow{\Omega}|=\pi$. Reciprocally, the presence of a kink reveals
the presence of dislocations attached to the line. The above picture of a kink
says nothing about the nature (edge or screw) of the attached dislocations, and
the way they relax and disperse through space about the disclination line. The
line flexibility, \textit{i.e.} the main property at work when the medium is
deformed, elastically or by flow, takes its origin here, in this interplay of
the
disclination line with dislocations.\\

A kink can be infinitesimally small;
\begin{equation}
\overrightarrow{db}_{disl}=2 \sin \displaystyle \frac
{\Omega}{2} \overrightarrow{t} \times \overrightarrow{ds},
\end{equation}
where $\overrightarrow{ds}$ is an infinitesimal element along
the line \cite{friedel1}. A density of infinitesimally small
kinks modifies the curvature of the line.  A dislocation
attached to an infinitesimally small kink has an infinitesimally
small Burgers vector; a dislocation attached to a
finite kink may have a finite Burgers vector, as we see now.\\

\subsection{Kinks in a SmA}

Let us now consider in more detail the geometry of the attachment of
dislocations to a focal line in a FCD. We first state some general properties,
and then consider separately the case of the ellipse and the case of the
hyperbola. Again, the dislocations emanating from the kink have to belong to
one of the two following categories: they are either dislocations with
infinitesimal Burgers vectors whose directions are parallel to the layer
dislocations of the layer stacking, or with Burgers vectors
$|\overrightarrow{b}_{disl}|=n d_0$ perpendicular to the layer (these are the
usual SmA quantified dislocations).  Note that in both cases the Burgers
vectors are translation symmetry vectors; they are perfect Burgers vectors in
the sense of the Volterra process. We consider them successively.

\textit{Infinitesimal Burgers vectors} relate to dislocation
densities that relax by the effect of viscosity; they affect the
curvature of the layers and consequently, as alluded just above,
they also affect their thickness, since the layers have to keep
in contact.  We shall not expatiate on such defects, which are
not relevant to our subject. Just note that the theory has been
developed for solids since long; see \cite{kroener} for a
general review. An essential point worth emphasizing is that a
continuous density of infinitesimal dislocations can be attended
by a strainless, elastically relaxed, state. In our case, this
would correspond to a state where the layers keep parallel.
Continuous dislocation with Burgers vectors parallel to the
layers do not introduce any kind of singularity of the SmA order
parameter. Eq. 4 indicates that the related kink
$\overrightarrow{ds}$ and that $\overrightarrow{t}$ are both
perpendicular to $\overrightarrow{db}_{disl}$, which condition
does not specify any special direction for
$d\overrightarrow{s}$.

\textit{Finite Burgers vectors}: this case is better represented by Eq. 3,
because the Burgers vector and the kink $\overrightarrow{AB}$ are finite.
$\overrightarrow{AB}$ and $\overrightarrow{t}$ have both to be in a plane
tangent to the local layer. To an elementary dislocation
$|\overrightarrow{b}_{disl}|=d_0$ corresponds an elementary kink. An elementary
kink is microscopic ($AB=2 d_0$); one can thus possibly have a density of such
elementary kinks, rendering the line curved when observed at a mesoscopic
scale. This does not exclude the possibility that infinitesimally small
dislocations are attached to finite kinks. Simple as they look, the application
of these
criteria requires however some care.\\

\subsection{Quantified Burgers vectors attached to an ellipse.}

Fig.4 is a schematic view of the properties of an ellipse,
belonging to an ideal FCD, which are in relation to its
$k=\displaystyle \frac{1}{2}$ disclination character.  The layer
geometry is different inside and outside the ellipse.
\underline{\textit{Inside}}, the Dupin cyclide layers intersect
the plane of the ellipse \textit{perpendicularly}.
\underline{\textit{Outside}}, the layers are planar and
\textit{perpendicular} to the asymptotic directions, as more
detailed below. The change of geometry between the inside and
the outside is achieved by a rotation of the layers about the
local rotation vector $\overrightarrow{\Omega}=\Omega
\overrightarrow{t}$; $\overrightarrow{\Omega}$ is parallel to
the layers (inside and outside) and is along the intersection of
the layers with the plane of the ellipse, inside.

The layer at M (M being a running point on the ellipse) is
indeed folded inside about the local $\overrightarrow{t}$
direction, is singular at M (it is a conical point), and extends
outside along a fold made of two half planes symmetrical with
respect the ellipse plane, each perpendicular to one or the
other of the two asymptotic directions of the confocal
hyperbola, and thereby making an angle about a direction
parallel to the minor axis of the ellipse (see
\cite{klemanbook}, chapter 10). The ellipse plane outside the
ellipse is therefore a tilt boundary of misorientation angle
$\omega$, which can be accommodated by edge dislocations of
Burgers vectors multiple of $d_0$, perpendicular to the plane of
the tilt boundary, \textit{i.e.} the plane of the ellipse
outside. There is one such dislocation
$|\overrightarrow{b}_{disl}|=2d_{0}$ per layer counted inside
the ellipse.\\

The same result can be obtained by using Eq. 4. Let us
parameterize the ellipse in polar coordinates with the origin at
the physical focus, Fig.6.\\

\begin{equation}
r=\displaystyle \frac{p}{1+e \cos \phi},
\end{equation}
where $p=\displaystyle \frac{b^2}{a}$ and $\phi$ is the polar
angle. One then finds that the $k=\displaystyle \frac{1}{2}$
ellipse disclination is partially of twist character, with an
attached Burgers vector

\begin{equation}
db_{disl} = 2 dr,
\end{equation}
The total Burgers vector attached to the ellipse is
$\displaystyle \int_{\phi=0}^{\phi=\pi} db_{disl}=4c$, as
indicated above. If one takes $dr=d_0$, - an approximation which
makes sense (up to second order), since $d_0$ is so small
compared to the size $a$ of the ellipse - it is visible that the
points $M\{{r, \phi}\}$ and $N\{{r + dr, \phi+d\phi}\}$ are on
two parallel smectic layers at a distance $d_0$. Notice that the
density of dislocations is constant if measured along the major
axis: $\displaystyle \frac{db_{disl}}{dx} =-2e$. There are no
dislocations attached to the singular circle of a toric FCD, as
the eccentricity e vanishes. An ellipse can be
thought of as a circle kinked at the layer scale.\\

\section{KINKED FOCAL CONIC DOMAINS}

\subsection{Frequent geometries for a kinked ellipse.}

The kinking of the ellipse takes different geometries, whether
the dislocations at stake are located  outside (where
$\overrightarrow{b_{disl}}$ is perpendicular to the plane) or
inside (where $\overrightarrow{b_{disl}}$ is in the plane) the
FCD.

\textit{Outside the FCD}: the Mouse (Fig.7). If the dislocation
lines attached to the ellipse disperse \textit{away outside} the
focal conic domain, \textit{i.e.} in a region of space where the
layers are in the plane of the ellipse; $\overrightarrow{t}$,
which varies in direction all along the ellipse, is in this
plane. Applying Eq. 3, it appears that the kinks have to be in
the plane of the ellipse. This configuration has been observed,
in a situation where the kinks are so small and have such a high
density that the kinked ellipse appears to be continuous, but
its shape departs considerably from a 'perfect' ellipse; it is
smoothly distorted by the in-plane kinks: we call it a 'Mouse'
(Fig.7a). Fig.7b provides a model for such kinks, (which always
go by pairs), drawn here at a scale which has no relation with
the real scale. The photograph of Fig.7a is taken from the rim
of a free standing film, in a region where the thickness $d$ of
the film is quickly changing, and the wedge angle $\omega$
between the opposite free boundaries varies monotonically. The
anchoring conditions are homeotropic; there is therefore a tilt
boundary in the mid-plane of the film, but with a variable
misorientation angle. The Mouse is in this mid-plane; the extra
dislocations attached to the kinks (edge dislocations in the
mid-plane) relax the variation of $\omega$ by contributing to
the modification of the density $\displaystyle
\frac{db_{disl}}{ds}$ of dislocations in this plane; see [2] for
a more detailed account.

\textit{Inside the FCD}: the Giraffe (Fig.8). The layers rotate
about $\overrightarrow{t}$ by an angle of $\pi$; hence they
become perpendicular to the plane of the ellipse,
\textit{inside} the FCD. Therefore the dislocations that
disperse away inside are attached to kinks that are
perpendicular to the plane of the ellipse, on average.  A pair
of elementary kinks (not at scale at all in the figure),
symmetric with respect to the major axis, can be linked by a
unique dislocation (Fig.8b). Our observations (Fig.8a) indicate
the existence of another mode of kinking, with screw
dislocations joining the kink (of macroscopic size) to the
hyperbola; another kink should then exist on the hyperbola, but
is not visible. We call such a departure from the perfect
ellipse, distorted by off-plane kinks, a 'Giraffe'. One can
eventually imagine elementary kinks of the sort in discussion,
all of the same sign, having a high density on the ellipse and
continuously tilting its plane. Such tilted ellipses have been
observed in 8CB and 9CB \cite{meyer1}. The situation observed in
Fig.8a results from the presence of a quasi planar
\textit{pretilted} anchoring. A unique direction of pretilt is
in conflict with the presence of an entire ellipse parallel to
the boundary in its close vicinity; hence opposite displacements
of different parts of the ellipse along the vertical direction,
to the point that one part gets off the boundary, and is
virtual; see \cite{yuriy} for a more detailed account of this
geometry and other geometries implying different kink types.
Fig.8c illustrates a double-kinked ellipse of a Giraffe type
observed from the side in a thick ($\approx200\mu m$) 8OCB
sample.

\subsection{On the origin of deviations from Darboux's theorem}

The just alluded kinking processes can bring large deviations to
Darboux's law; reciprocally it is clear that the deviations from
Darboux's law mean a modification of the shape of the ideal FCD
conics, i.e. the presence of kinks (at the scale of the layers,
since they are not visible with the optical microscope) and of
their attached dislocations.  These dislocations necessarily
disperse through the medium, outside and/or inside the FCD.
Infinitesimal dislocations, if alone, would result, as stated
above, in an extra curvature of the layers; two cases arise:
either the deformed layers keep parallel, hence the layer normal
keep straight, and one gets eventually a new ideal FCD, or there
is a deviation to straightness of the layer normals, and
consequently a layer thickness variation (this case falls within
the province of the Kroener's dislocation densities
\cite{kroener}), i.e. a process of high energy if not relaxed, at
least in part, by finite edge dislocations. It suffices then to
consider only those latter. The edge components of the attached
dislocations that are dispersed inside the FCD break the
parallelism of the inside layers. The congruence of the layer
normals is thus no longer a set of straight lines. This is another
way of explaining the variation to Darboux's theorem. This could
have been stated from the start: \textit{edge dislocation
densities break Darboux's theorem, because they break the layer
parallelism}. But this statement comprehends deviations to
Darboux's theorem that are more general than those where the focal
manifolds are degenerate to lines. The focal manifolds of a
congruence of curved normals are generically 2D surfaces, not
lines. We see that the fact that these surfaces are degenerate
into lines comes from the fact that the dislocations in question
are \textit{attached} to the original focal lines. To conclude,
the occurrence of deviations to Darboux's theorem for a set of
focal \textit{lines} means that the conics are (densely) kinked
and dislocations attached to those kinks.\\

\subsection{The kinked (split) hyperbola.}

The shape of the layers is cylindrical about the central zone of the hyperbola,
near its apex (which is also the physical focus of the ellipse). But the layers
are practically perpendicular to the hyperbola at a distance of order $a$ to
the plane of the ellipse; the wedge disclination smoothly vanishes far from the
ellipse plane.  In between, the layers display cusps, the lesser pronounced the
more distant from the ellipse.  Hyperbolae are lines of easy coalescence of
screw dislocations, as observed long ago \cite{williams2}.

The presence of kinks on the hyperbola is a delicate matter; because it is a
$k=1$ wedge disclination ($\Omega= 2\pi$), Eq. 3 and 4 do not apply directly.
A way of solving the question is to consider that the line is made of two
$k=\displaystyle \frac{1}{2}$ lines, indicating that dislocations with total
Burgers vectors twice as large can attach to a kink of the same size as in the
$k=\displaystyle \frac{1}{2}$ case.

Another situation is worth considering. In fragmented focal
domains of the type represented Fig.9b (called
\textit{fragmented} domains), the hyperbola belongs to the
boundary of the domain.  It is then no longer a $k=1$
disclination but a $k=\displaystyle \frac{1}{2}$ disclination,
as if it were split all along its length. Such an object, noted
sFCD for short, and already recognized by G. Friedel
\cite{friedelb}, is easily obtained in a confined sample. A sFCD
is bound by a segment of the ellipse and by a segment of the
hyperbola, and four fragments of cones of revolution. Thus both
segments are $k=\displaystyle \frac{1}{2}$ disclination line
segments.

As a consequence, sFCDs are generally aligned, attached by the ends of the
disclination segments, such attachments being required by the conservation of
the disclination strength.  But observe that a hyperbola H (resp. an ellipse E)
can be attached indifferently either to another H (resp. an E) or to an E
(resp. a H).

One can imagine that the ellipse $E_1$ of a $FCD_1$ is attached
to $H_2$ of a $FCD_2$, while the hyperbola $H_1$ of the $FCD_1$
is attached to $E_2$ of the $FCD_2$. Such a set of line segments
attached by their extremities is topologically equivalent to a
double helix.  This geometry, with sequences of the ...HEHEH...
type, was observed long ago by C. E. Williams \cite{williams1}
at the N $\longrightarrow$ Sm transition; it is at the origin of
helical giant screw dislocations.

Let us also mention the observation, also reported in
\cite{meyer1}, of a mobile kink (several microns long)
perpendicular to the $k=\displaystyle \frac{1}{2}$ hyperbola of a
sFCD, moving in the direction of the physical focus, but nucleated
far from it, at a distance large compared to $a$. There is no
doubt that dislocations, dragged along the hyperbola, are attached
to this mobile kink; their Burgers vectors, that are perpendicular
to the layers, are practically parallel to the asymptotic
direction of the hyperbola, at a distance from the ellipse plane,
which indicates that they are of screw character. This might be an
indication of a mechanism by which screw dislocations align along
a (split) hyperbola.\\

\subsection{Focal Conic Domains at the Sm $\longrightarrow N$ transition}

FCD's that are immersed in the bulk (they are of the type
represented Fig.9a, and generally gather into tilt boundaries)
disappear rather suddenly about $0.5^{\circ}C$ before the
transition, by an instability mechanism  which certainly implies
a sudden multiplication of dislocations. The capture of free
edge dislocations by the ellipse modifies its geometric features
$e$ and $a$, Fig.10. Free dislocations of the same (resp.
opposite) sign as the dislocations attached to the ellipse, if
captured, would increase (resp. decrease) its size ($2a
\longrightarrow 2a+b_{disl}$), either at $e$ constant (then the
asymptotic directions stay constant), or not. Boundary
conditions play a dominant role in this relaxation process.
Notice that, after a possible increase in size, the ellipses
eventually always decrease in size when the temperature
increases, the smallest ellipses disappearing first. For the
ellipses belonging to a grain boundary, this implies that the
boundary area occupied by dislocations (the so-called residual
boundary) increases with temperature. This is in agreement with
the model developed in \cite{kleman1}, which relates the
residual boundary to the material constants; in particular a
decrease of the compression modulus $B$ must result in an
increase of the residual area. Another important issue here is
the existence of the instability. Both topics will be discussed
in more details in a forthcoming
publication.\\

\section{CONCLUSIONS}

This paper investigates from a theoretical point of view some
features of the FCD transformations that have been observed, in
the smectic phase, when approaching the nematic phase. These
very spectacular phenomena happen in a large temperature domain
($\Delta T=T_{AN}-T^* \approx$ half a degree in 8CB, which is
the chemical we used for quantitative observations; the other
compounds yield qualitatively equivalent results) in which it is
believed that the variations of the material constant $B$ are
large enough to allow significant variations of the dislocation
line energy and the multiplication of fresh dislocations.  At
the same time $K_1$ and also $\bar{K}$ (as we assume) do not
vary in comparable proportion, so that the energy of focal conic
domains is not appreciably changed.

We have tried to discuss the general principles at the origin of
these transformations that are due to the direct interaction
between FCDs and finite Burgers vector dislocations. There is no
doubt that infinitesimally small Burgers vector dislocations are
also playing a role, in particular in the phenomena of viscous
relaxation \cite{klemanbook,friedel3}, but this is not
discussed. The general principles that we advance are
geometrical and topological in essence.  The mechanisms that
obey these principles seem to be plenty, depending in particular
on the boundary conditions and the precise FCD texture.  The
examples we have given are few, and are chosen for the sake of
illustration. A description of several more observed
transformations, interpreted in the same terms, will be given
somewhere else.

The SmA $\longrightarrow$ N transition is one of the most
debated liquid crystal phase transitions
\cite{degennes,tonera,tonerb,helfrich}. This is not the place to
enter into the detail of this debate, inasmuch as our results,
even if they stress the importance of defect interplays in the
critical region, are not directly related to the very proximity
of the transition, which has been examined by several authors
with great accuracy (\textit{e.g.} \cite{yethiraj}).

The question which is at stake is rather why the interactions
occur at temperatures definitively lower than $T_{AN}$. We
interpret this phenomenon as an instability for the multiplication
of dislocations, much akin to a Kosterlitz-Thouless transition
\cite{kosterlitz} under temperature, but also under stress (the
boundary conditions), of the sort proposed in \cite{khanta} for a
completely different type of transition.  More details on the
quantitative nature of the transition will be given in a
forthcoming publication.

\section{APPENDIX}

We envision a curved disclination line $\cal L$, carrying a rotation vector
$\overrightarrow{\Omega}$ constant in length and in direction.  Let $P$ be a
point on the cut surface bound by $\cal L$.

We first assume that $\overrightarrow{\Omega}$ is attached to
some well-defined point $O$ (Fig.11). The relative displacement
of the
two lips of the cut surface at $P$ is:\\
\begin{equation}
\overrightarrow{d}_{P}(O)=\overrightarrow{\Omega} \times \overrightarrow{OP}
\end{equation}
which is large on the line $\cal L$ if $P$ is taken at some point $M$ on $\cal
L$. Consequently in the generic case $\cal L$(0, $\Omega)$ has a very large
core singularity, thus large accompanying stresses and a large core energy. On
the other hand the cut surface displacement vanishes at $M$ if
$\overrightarrow{\Omega}$ is attached to $\cal L$ at $M$, but then it does not
vanish at $N=M+\overrightarrow{dM}$. There is still a large core singularity
along $\cal L$, except at $M$. The Volterra process, when applied in its
standard form, does not provide a solution to the construction of a curved
disclination with well relaxed
stresses.

An extended conception of the Volterra process solves the problem. Assume that
there is a copy of the rotation vector $\overrightarrow{\Omega}$ attached to
all the points of $\cal L$, and consider the effect of such on a point $P$
belonging to the cut surface of all these $\overrightarrow{\Omega}'s$. We have,
for each other $M$ belonging to $\cal L$, another value of the relative
displacement of the lips of the cut
surface:\\
\begin{equation}
\overrightarrow{d}_{P}(M)=\overrightarrow{\Omega} \times \overrightarrow{MP}
\end{equation}
This difficulty is easily solved by the introduction of a set of
\textit{infinitesimal} dislocations attached to the disclination
line all along (Fig.12). Let $M$ and $N=M+\overrightarrow{dM}$
be two infinitesimally close
points on $\cal L$. We have:\\
\begin{equation}
d_\textbf
{P}(\overrightarrow{M}+d\overrightarrow{M})-d_\textbf{P}(\overrightarrow{M})=\overrightarrow{\Omega}
\times d\overrightarrow{M}
\end{equation}
which is independent of $P$. The quantity $d\overrightarrow{b(M)}
=\overrightarrow{\Omega} \times \overrightarrow{dM}$ is the infinitesimal
Burgers vector of the infinitesimal dislocation attached to $\cal L$ at point
$M$ \cite{friedel1}.

The above equations are established for a small angle of
rotation vector $|\overrightarrow{\Omega}|$.  In the general
case $\Omega$ has to be replaced by $\displaystyle \frac{1}{2}
\sin \frac{\overrightarrow{\Omega}\overrightarrow{t}}{2}$, where
$\overrightarrow{\Omega}=\Omega \overrightarrow{t}$.

\acknowledgements We acknowledge fruitful discussions with V. Dmitrienko and J.
Friedel. We are grateful to J.-F. Blach for providing us with glass plates
treated for special anchoring conditions.

\begin{center}
Figure Captions
\end{center}

Fig.1: a) Complete FCD with negative Gaussian curvature Dupin
cyclides, sitting inside cylinders of revolution meeting on the
ellipse.  The cyclides cross the ellipse plane at right angles;
their intersections with the ellipse and the hyperbola, when
they exist, are conical points. b) Dupin cyclides fragments with
positive and negative Gaussian curvature, so chosen that the
ellipse is still singular but the hyperbola has no physical
realization.  An opposite situation (ellipse with no physical
realization, hyperbola still singular) is illustrated in
\cite{klemanbook}.

Fig.2: In an ideal FCD the ellipse and the hyperbola project
orthogonally along two conics which intersect at right angles,
as observed (long side of photographs $\approx 200\mu m$): a)
8OCB thick ($\approx 200\mu m$) sample annealed during about 48
hours deeply in the SmA phase ($7 ^{o}C$ below the transition
from the nematic phase) between two untreated glass substrates;
Darboux's theorem obeyed; the set of the FCD's with parallel
hyperbolae asymptotes form a grain boundary of the type
schematically shown in Fig.3; b)  8CB ($0.5 ^{o}C$ below the
transition to the nematic phase); Darboux's theorem disobeyed as
demonstrated for the lower photograph: solid lines are tangents
to the disclinations and the dashed lines are orthogonal; a very
visible deviation from the Darboux's theorem is encircled on the
upper photograph.

Fig.3: Tilt boundary split into FCDs. a) schematic, adapted from
\cite{klemanbook}; b) 8CB, polarized light microscopy
observation; the tilt boundary is seen edge-on; the edge of the
photograph $\approx 100\mu m$ long.

Fig.4: F, the physical focus, is the center of the (circular)
intersections of the layers with the plane of the ellipse,
inside the ellipse; $\overrightarrow{t}$ is a unit vector along
the local rotation vector; the $k=\displaystyle \frac{1}{2}$
disclination ellipse is of mixed (twist-wedge) character all
along, except at the ends of the major axis, where it is wedge.

Fig.5: Kink on a wedge disclination line, see text.

Fig.6: The ellipse in polar coordinates.  The radius of
curvature of the circle centered in the focus F and tangent to
the apex is $a-c$, which is smaller than the radius of curvature
$\displaystyle \frac{b^2}{a}$ of the ellipse at the apex. This
circle is thus entirely inside the ellipse.  All the circles and
the arcs of circles of the figure are centered in F. They figure
intersections of the smectic layers with the plane of the
ellipse.

Fig.7: Double kinks with a dislocation \textit{outside} the FCD;
a) Mouse patterns in 8CB, free standing film, rim region; the
thickness decreases downwards; long side of the photograph
 $\approx 200\mu m$); b) model.

Fig.8: Views of a double kink with a dislocation \textit{inside}
the FCD (long side of photographs $\approx 200\mu m$): a)
Giraffe patterns in 8CB , demonstrating that the ellipses are
divided into two parts  not located at the same level as
depicted in the model below, the screw dislocations attached to
the kinks are visible; b) model of a double kink linked by a
unique dislocation located inside the FCD; c) - a double kinked
ellipse (kinks are shown by arrows) of the Giraffe type observed
from the side (8OCB in a gap of the thickness $\approx 200\mu m$
between two untreated glass substrates).

Fig.9: Incomplete FCD's. a) FCD bound by two cones of revolution
meeting on the ellipse, with apices at the terminations of the
hyperbola segment; b) A hyperbola-split fragmented FCD (sFCD).
The sFCD is bound by i) two fragments of cones of revolution
with apices at the terminations of the hyperbola segment and
limited to the ellipse segment, ii) two fragments of cones of
revolution with apices at the terminations of the ellipse
segment and limited to the hyperbola segment. The director field
on the boundaries is indicated, not the cyclide intersections.

Fig.10: Edge dislocations mobile in the plane of the ellpse and
attaching to it. The consecutive relaxation process modifies the
FCD according to the boundary conditions.

Fig.11: The classic Volterra process for a rotation vector
$\overrightarrow{\Omega}$ attached to $O$. At a point $P$ on the
cut surface, the lips of the cut surface suffer a relative
displacement $\overrightarrow{d}_P(O) = \overrightarrow{\Omega}
\times \overrightarrow{OP}$.

Fig.12: The extended Volterra process for a rotation vector
$\overrightarrow{\Omega}$ attached locally to each point on
$\cal L$. Infinitesimal dislocations are attached all along the
disclination line.


\begin{references}

\bibitem{meyer1}
C. Meyer and M. Kleman, ILCC2004, Mol. Cryst. and Liq. Cryst.,
under press.

\bibitem{yuriy}
Yu. A. Nastishin, C. Meyer and M. Kleman, in preparation.

\bibitem{degennes}
P. G. De Gennes and J. Prost, \textit{The Physics of Liquid Crystals}, $2^{nd}$
edition, Clarendon Press (1993).

\bibitem{benzekri}M. Benzekri, T. Claverie, J.-P. Marcerou,
and J.-C. Rouillon, Phys. Rev. Lett. 68, 16 (1992).

\bibitem{beaubois} F. Beaubois, T. Claverie, J.-P. Marcerou, J.-C. Rouillon,
and H.-T. Nguyen, C.W. Garland and H. Haga, Phys. Rev. E \textbf{56}, 5 (1997).

\bibitem{yethiraj} A. Yethiraj, J. Bechhoefer, Phys. Rev. Lett. \textbf{84}, 16 (2000).

\bibitem{barbero}
G. Barbero and V. M. Pergamenshchik, Phys. Rev. E \textbf{66} 051706 (2002).

\bibitem{friedela} G. Friedel et F. Grandjean, Bull. Soc. Fr. Min\'{e}r. \textbf{33}, 192, 409 (1910).

\bibitem{friedelb}G. Friedel, Ann. Phys. (Paris) \textbf{18}, 273 (1922).

\bibitem{klemanbook}
M. Kleman and O. D. Lavrentovich, \textit {Soft Matter Physics.
An Introduction}, Springer N.Y. (2003).

\bibitem{maurice}
M. Kleman and O. D. Lavrentovich, Phys. Rev. E\textbf{61}, 1574 (2000).

\bibitem{boltenhagen1}
P. Boltenhagen, O. Lavrentovich and M. Kl\'{e}man, Phys. Rev. A46, 1743(1992).

\bibitem{allaina} M. Allain and J.-M. diMeglio, Mol. Cryst. Liq. Cryst. \textbf{124}, 115
(1985), Europhys. Lett. \textbf{2}, 597 (1986).

\bibitem{allainb} M. Allain and M. Kleman, J. de Physique \textbf{48}, 1799 (1987).

\bibitem{dhez}
O. Dhez, S. K\"{o}nig, D. Roux, F. Nallet and O. Diat, Eur. Phys. J. E\textbf{3},
377 (2000).

\bibitem{williams1} C.E. Williams, Philos. Mag. \textbf{32}, 313 (1975).

\bibitem{kln}
M. Kleman, O. D. Lavrentovich and Yu. A. Nastishin, in Dislocations in Solids,
F. R. N. Nabarro and J. P. Hirth, eds, North-Holland, Amsterdam vol.
\textbf{12}, 147 (2004).

\bibitem{darboux} G. Darboux, \textit{Le\c{c}ons sur la th\'{e}orie g\'{e}n\'{e}rale des surfaces}, $2^{\grave{e}me}$ partie,
Gauthier-Villars, Paris (1914).

\bibitem{kleman1} M. Kleman and O. D. Lavrentovich, Eur. Phys. J. E\textbf{2}, 47 (2000).

\bibitem{bourdon} L. Bourdon, M. Kl\'{e}man and J. Sommeria, J. de Physique \textbf{43}, 77 (1982).

\bibitem{boltenhagen2} P. Boltenhagen, O. D. Lavrentovich and M. Kl\'{e}man, J. Phys. II France
\textbf{1}, 1233 (1991).

\bibitem{frank} F. C. Frank, Discuss. Faraday Soc. \textbf{25}, 19 (1958).

\bibitem{friedel1} J. Friedel and M. Kl\'{e}man, J. de Phys. \textbf{30}, C4:43 (1969).

\bibitem{kroener} E. Kroener, in Physics of Defects, Les Houches 1980 Session XXXV (eds. R.
Balian, M. Kl\'{e}man and J.P. Poirier), North-Holland, Amsterdam, (1981).

\bibitem{williams2} C. E. Williams and M. Kl\'{e}man, Philos. Mag. \textbf{33}, 213 (1976).

\bibitem{friedel3} J. Friedel, private communication.

\bibitem{tonera} J. Toner and D. Nelson, Phys. Rev. B\textbf{24}, 363 (1981.

\bibitem{tonerb} J. Toner, Phys. Rev. B\textbf{26}, 1 (1982).

\bibitem{helfrich} W. Helfrich, J. de Physique \textbf{39}, 1199 (1978).

\bibitem{kosterlitz} J.M. Kosterlitz and D. J. Thouless, J. Phys. C: Solid State Phys., \textbf{6},
1181 (1973).

\bibitem{khanta} M. Khanta, D. P. Pope and V. Vitek, Phys. Rev. Lett. \textbf{73}, 684 (1994).


\end{references}
\end{document}